\documentclass[preprint,tightenlines,eqsecnum,floats,aps,amsmath,amssymb,nofootinbib,prd,showpacs]{revtex4}
\usepackage{amssymb}
\usepackage{amsmath}
\usepackage{graphicx}
\usepackage[normalem]{ulem}
\usepackage[dvips]{color}

\begin{document}

\title{ Universal behavior of compact star based upon gravitational binding energy }

\author{Rongrong Jiang}
\affiliation{School of Physics and Optoelectronics, South China University of
Technology, Guangzhou 510641, P.R. China}
\author{ Dehua Wen\footnote{Corresponding author. wendehua@scut.edu.cn}}
\affiliation{School of Physics and Optoelectronics, South China University of
Technology, Guangzhou 510641, P.R. China}
\author{Houyuan Chen}
\affiliation{School of Physics and Optoelectronics, South China University of
Technology, Guangzhou 510641, P.R. China}

\date{\today}

\begin{abstract}
 Similar to the compactness parameter ($\beta=M/R$),  the gravitational binding energy (GBE) is also a characteristic parameter which can reflect the internal structure of a neutron star and thus can be used to expressing the universal relations.
 Scaling by the stellar mass, this investigation demonstrates a perfect universal relation between the GBE and the moment of inertia, where both of the normal neutron stars and the quark stars satisfy the same universal relation.
  Moreover, a fine empirical relation between the GBE and the tidal deformability  is proposed, where the difference of the relations can be used to distinguish wether  a pulsar is a normal neutron star or a quark star if the stellar mass and the tidal deformability  can be observed or estimated rather accurately.
 These universal relations provide a potential way to estimate the GBE if the stellar mass and the moment of inertia/the tidal deformability are precisely measured.

\end{abstract}


\maketitle

\section{Introduction}

The equation of state (EOS),  which is basically seen as the density-pressure relation, virtually governs most macroscopic properties of neutron stars, such as their mass ($M$), radius ($R$), moment of inertia ($I$), quadrupole moment ($Q$) and tidal deformability ($\Lambda$). The majority composition of a neutron star is currently considered as kind of cold and ultra-dense nucleonic fluid. Presently, people can not well understand the EOS of nuclear matter beyond saturation density only through the  terrestrial experiments \cite{Li2008,Ozel2016,Feng2018}.  Fortunately, the astronomical observations on the pulsars may provide the unknown segments of the EOS \cite{Li2019EPJA}.
Nowadays, more and more useful observations of neutron stars have been accumulated, especially the  discovery of massive neutron stars in recent years \cite{Demorest2010,Antoniadis2013,Cromartie2019} and the gravitational wave radiation detection of the merging of binary neutron stars \cite{LIGO2017}, which have provided effective constraints on the EOSs \cite{Miller2015,Ozel2016,Zhang2018,Zhang2019,Margalit2017,Rezzolla2018,Ruiz2018,LIGO2018,Annala2018,Most2018,De2018,Chakravarti2019,Chen2019CPC,Zhang2019PRC}.
It has been an important thread to use the neutron star observations to inversely study the EOS.
 For example, Bayesian inference was adopted frequently to explore and contrast the parameters of different EOSs models \cite{LIGO2018,Most2018,Annala2018,Greif2019}.

Among the numerous studies on this issue, to find universal relations (independent from the EOS)  between the properties of neutron star is a practicable and effective method.
In the last twenty years or so, a great deal of research work has been done on these universal relations. For example, the universal relations of the quasi-normal modes in neutron star provide an effective method in studying the  frequency and damping time of the oscillation, see, e.g., refs. \cite{Andersson1996,Tsui2005,Lau2010,Blazquez2014,Stergioulas2018,Wen2019fmode}.
In the past few years, a new type of universal relations between the properties, including the moment of inertia, the Love numbers and the normalized quadrupole moment, namely I-Love-Q,  was established \cite{Yagi2013a,Yagi2013b} and  thoroughly investigated \cite{Doneva2014,Pappas2014,Chakrabarti2014,Haskelli2014,Majumder2015,Carson2019,Rosofsky2019,Wen2019fmode}.
These relations  are very useful as they provide a direct link to  $I$, Love number and $Q$,  and if the stellar mass  and one of the three parameters of a neutron star is observed, and then the other two properties  can be determined.
Someone may ask that though the universal relation is independent from the EOS, how can we extract the information of the EOS? In fact, as only a few of the global properties can be observed accurately,  for those properties which are difficult to be observed precisely, such as the radius, the moment of inertia, they can be obtained through combining the universal relations and the accurately observed property. And then we can further determine the EOS as the derived global properties of neutron star depend strongly on the EOS \cite{Lau2010,Wen2019fmode}.

In the research of finding and constructing the universal relations between the properties of neutron star, the compactness parameter ($\beta=M/R$) and the  moment of inertia  are   considered to be the two key parameters. Why are these two quantities so important  in the universal relations? As pointed out in  refs. \cite{Tsui2005}, most of the properties of neutron stars, such as the  the Love number, the normalized quadrupole moment and the parameters of the quasi-normal modes are related to the stellar mass distribution $m(r)$, and the mass distribution can be solely characterized by the compactness and be independent of the stellar mass $M$ \cite{Tolman1939,Lattimer2001}. The moment of inertia is believed to be a better parameter to describe the global mass distribution and thus leads to an improved universal behavior \cite{Lau2010}.

Could there be better parameters  to construct the universal relations? Lattimer \textit{et al}. proposed the existence of approximate universality between binding energy and compactness of neutron star \cite{Lattimer2001,Prakash1997}. But this universality is too rough to be used as an accurate relation to determine the properties of neutron star. In fact, the problem comes from the binding energy. The binding energy in their universal relation is the total binding energy, including both of the gravitational binding energy and the nuclear binding energy. Obviously, the nuclear binding energy does not include the mass distribution information.
Inspired by that the dimensionless gravitational binding energy in Newtonian gravity of a uniform sphere can be expressed as $\frac{E_{g}}{M}=\frac{3}{5}\frac{M}{R}$ ($E_{g}$ is the gravitational binding energy), linked directly to the compactness, we believe that the gravitational binding energy would be a better parameter to build the universal relations between the properties of neutron star.
 Actually, this work discovers two interesting EOS-insensitive universal relations based upon gravitational binding energy. There are $I$-$E_{g}$ and $\Lambda$-$E_{g}$, respectively.
  In this work, we will present and discuss these universal relations in details.

The paper is organized as follows. In Sec. \ref{BNCNS}, the binding energy and its measurement are briefly reviewed. In Sec. \ref{Equation}, the parametric EOS of neutron-rich dense matter and  a few of other EOS models are  concisely introduced. In Sec. \ref{URRBE}, we present the
 universal relations between the gravitational binding energy and the global properties of neutron stars in details.  A brief summary is given at the end.

Unless otherwise noted, we use geometrical unit (G = c = 1).

\section{Review of the binding energy in neutron star} \label{BNCNS}

The stellar mass measured by observer at infinity is defined as the gravitational mass. For a non-rotating neutron star, the gravitational mass can be calculated by \cite{Oppenheimer1939,Glendenning2000,Hartle2003}
\begin{equation}
M=\int_{0}^{R} \rho(r)4\pi r^2\, dr,
\end{equation}
where $\rho (r)$ is the mass density at radius $r$, and $R$ is the surface radius of a neutron star.
As the proper volume element in Schwarzschild metric is written as \cite{Glendenning2000,Hartle2003,Cameron1959,Bagchi2011}
\begin{equation}
dV=\sqrt{-g}\, d^4x=4\pi r^2[1-\frac{2m(r)}{r}]^{-\frac{1}{2}}\, dr,
\end{equation}
the total baryon number of a non-rotating neutron star can be obtained through \cite{Glendenning2000,Hartle2003}
\begin{equation}
A=\int_{0}^{R} n(r) dV=\int_{0}^{R}n(r) 4\pi r^2[1-\frac{2m(r)}{r}]^{-\frac{1}{2}}\, dr,
\end{equation}
where $n(r)$ is the number density of the baryon, and $m(r)$ is the mass within the radius $r$.

The baryon mass (in some references named as the rest mass) of the neutron star is defined as
\begin{equation}
M_b=A m_b,
\end{equation}
where $m_b$ is the mass of a baryon, which takes a value of 939 MeV for both of the neutron and proton. The difference
\begin{equation}
E_t=M-M_b
\end{equation}
is defined as the total binding energy of a neutron star \cite{Prakash1997,Lattimer2001,Lattimer1989,Lattimer2007}, which contains the total energy by assembling all of the baryons from infinity to form a stable neutron star, that is, contains both the gravitational binding energy and the nuclear binding energy. It is worth noting that in order to distinguish the attractive potential energy and the repulsive potential energy from the sign of the data, here we follow the rules that the negative value represents the attractive potential energy while the positive value represents the repulsive potential energy.

Through defining a proper mass of a neutron star  \cite{Cameron1959,Bagchi2011}
\begin{equation}
M_p=\int_{0}^{R} \rho(r)4\pi r^2[1-\frac{2m(r)}{r}]^{-\frac{1}{2}}\, dr,
\end{equation}
one can obtain the gravitational binding energy as
\begin{equation}
E_g=M-M_p.
\end{equation}
Obviously, the gravitational binding energy $E_g$ does not include the nuclear binding energy $E_n$. Through the definition of total binding energy and gravitational binding energy, the nuclear binding energy can be calculated through
\begin{equation}
E_n=E_t-E_g=M_p-M_b.
\end{equation}
To a normal neutron star, the total nuclear binding energy is positive, which means that the nuclear interaction between most of the nucleons is repulsive in neutron star. As a comparison, the nuclear binding energy per particle of isospin symmetric nuclear matter at saturation is about -16 MeV \cite{Lattimer2007}.

It has long been recognized that the binding energy of a neutron star can be measured through the supernova neutrino, as the supernova energy is mainly released in the form of neutrinos (at least 99\% \cite{Prakash1997}) and the released energy of the supernova can be approximated to the binding energy of the neutron star \cite{Prakash1997,Lattimer2001,Lattimer1989,Lattimer2007,Rosso2017}. As has been pointed out by Lattimer and Prakash \cite{Lattimer2001},  the released energy in a supernova explosion is from the collapse of a white-dwarf-like iron core but not from the free-baryons collapse from infinity, thus the measured binding energy through the supernova neutrino is not the total binding energy $E_t$, but the effective total binding energy $E_{et}=M-m_{eb}A$, where $m_{eb}$ is the effective mass of a baryon, takes a value of 930 MeV, corresponding to the mass of Fe$^{56}/56$.

The intermediate-mass neutron stars in the range of $1 \sim 1.5 M_{\odot}$ are expected to possess the total binding energy $E_t$ as high as $0.08 \sim 0.16 M_{\odot}$ \cite{Lattimer1989}. The formation of a massive neutron star would generally release greater energy than that of a less-massive one. The heaviest neutron star in the sky is therefore believed to get bound by enormous binding energy \cite{Prakash1997}. The measurements of neutrinos from SN1987A shows that the effective total binding energy is about $0.1 \sim 0.2 M_{\odot}$ for the neutron star with mass in  $1.14 \sim 1.55 M_{\odot}$ \cite{Hirata1987,Bionta1987}. Based on the analysis of the observation of $\gamma$-rays from $ ^{56}$Co and $ ^{57}$Co, Bethe and Brown obtained the baryon mass of the core left by the SN1987A as $M_b=1.733 \pm 0.024 M_{\odot}$ \cite{Bethe1995}.
Referring to the observation of SN1987A, we believe that the gravitational mass $M$, the total baryon number $A$, the baryon mass $M_b$ and the effective total binding energy $E_{et}$ (then the total binding energy $E_t$) can be measurable or be deducible through the detection of the supernova explosion in the future. With more observations on the supernova  in the future, the binding energy will have become an important character to tell the internal secrets of the compact neutron stars.

\section{EOSs  of neutron star and quark star models}\label{Equation}

 There are still great difficulties to extrapolate the current EOS into the density of neutron star core.  The predictions of the different EOS models often occur significant divergence at supra-saturation densities.
It is instructive to construct an EOS model not only minimizing the model dependence of EOS  but also containing all known constraints on the EOS. The   general parametric EOS model for neutron-rich nucleonic matter in the core is such an ideal model \cite{Zhang2018}.
In this work, we will employ this EOS model to describe  the neutron star core matter, which consists of protons, neutrons, electrons and muons at $\beta$-equilibrium. More details for this EOS model please refer to ref. \cite{Zhang2018} and the references therein.
For the outer crust and the inner crust, the BPS EOS \cite{Baym71} and the NV EOS \cite{Negele73} are adopted respectively.
By changing the EOS parameters of this parametric model, we can generate a huge number of EOSs for neutron star.
The parameters are essentially coherent with the terrestrial experiments on the nuclear physics. In addition, we rule out the EOSs that can not support a maximum mass greater than 2.01 M$_{\odot}$ or cannot meet the causal constraint.  Here we employ about 10,000 screened EOSs to investigate the universal relations.

For comparisons, we also employ 11 EOSs \cite{Zhang2018,Wen2019fmode} for normal or hybrid neutron stars constructed by microscopic nuclear many-body theories (marked as 11 microscopic EOSs in the following text and figures) and 3 EOSs  for the quark star models \cite{Fowler1981}. The 11 microscopic EOSs  are:
ALF2 of Alford et al. \cite{ALF2} for hybrid stars (nuclear + quark matter),  APR3 and APR4 of Akmal and Pandharipande \cite{AP34}, ENG of Engvik et al. \cite{ENG}, MPA1 of Muther, Prakash and Ainsworth \cite{MPA1}, SLy of Douchin and Haensel \cite{SLy}, WWF1 and WWF2 of Wiringa, Fiks and Fabrocini \cite{WFF12}, the QMFL40, QMFL60 and QMFL80  model  from the work of Zhu et al. \cite{QMF}. The three EOSs of quark stars are from the confined-density-dependent-mass (CDDM) model \cite{Fowler1981}. Similar to the work of ref. \cite{ALi2016},  here we also adopt the three typical EOSs (labelled as CIDDM, CDDM1, and CDDM2) as the quark star models.

 \section{Universal relations in terms of gravitational  binding energy} \label{URRBE}

  Based on a variety of equations of state, Lattimer et al. proposed that there exists a universal relation between the total binding energy and the stellar mass of the neutron star \cite{Lattimer1989}, namely
\begin{equation}
{|E_t|}\approx 0.084 (\frac{M}{M_{\odot}})^2 M_{\odot}.
\end{equation}

The formula is potentially applicable in determining the mass of neutron star  through the binding energy.
 In a later time, Lattimer et al. further proposed a relatively accurate universal relation of the binding energy as \cite{Lattimer2001}
\begin{equation}
\frac{|E_t|}{M}\approx \frac{(0.6 \pm 0.05)\beta}{(1-0.5\beta)},
\end{equation}
where $\beta=M/R$ is the compactness of a neutron star. Recently, Breu and Rezzolla gave a more precise universal relation between the mass and the binding energy through a quadratic polynomial function \cite{Breu2016}
\begin{equation}
\frac{|E_t|}{M}= d_{1}\beta+d_{2}\beta^{2},
\end{equation}
 where $d_{1}=6.19\times 10^{-1}$ and $d_{2}=1.359\times 10^{-1}$ .
 More interestingly, the moment of inertia was also found to link with the compactness \cite{Lattimer2001}. since both the total binding energy and the moment of inertia have universal relation with the compactness, it is natural to consider the relation between the total binding energy and the moment of inertia. Steiner et al. obtained such an  universal relation as \cite{Steiner2016}
\begin{equation}
\frac{|E_t|}{M}=0.0075+(1.96_{-0.05}^{+0.05})\bar{I}^{-1}-12.80\bar{I}^{-2}+72.00\bar{I}^{-3}-(160_{-20}^{+20})\bar{I}^{-4},
\end{equation}
where $\bar{I}=I/M^3$.

All the above  three universal relations are related to the total binding energy.
 From the Figs. 17, 18 and 24 of ref. \cite{Steiner2016}, it is shown that the total-binding-energy related universal relations are rather rough. As mentioned above, this is because the total binding energy includes both of the nuclear binding energy and the gravitational binding energy, and only the later  has the mass distribution information, which is just the essential internal cause of the  universal relations \cite{Tsui2005,Lau2010}.

 In order to make quantitative discussions on the  binding energy,   we firstly present the baryon mass, the proper mass, the total binding energy, the gravitational binding energy, the nuclear binding energy and their corresponding single nucleon energy for the canonical stars ($1.4~ M_{\odot}$) in Tab. \ref{tab:table1} and for the maximum-mass stars in Tab. \ref{tab:table2}.  From Tab. \ref{tab:table1}, it is easy to see that except for the hybrid star (ALF2) and the quark star (CDDM) models, all the canonical stars have similar baryon mass, proper mass, total  binding energy,   gravitational binding energy and nuclear binding energy. For these canonical neutron stars, their gravitational binding energy is about 1.3-1.6 times  the total binding energy, and their nuclear binding energy is positive.  For the hybrid star and quark stars, the nuclear binding energy is negative. Through Tab. \ref{tab:table2},
it is shown that the maximum-mass stars have much larger binding energies than their corresponding canonical neutron stars, and the average gravitational binding energy per nucleon ($E_g/A$) can be up to -325 MeV, while for the canonical neutron stars, the highest $E_g/A$ is about -155 MeV.

\begin{table}[htbp]
	\centering
	\caption{The binding energies of neutron stars and quark stars with a canonical gravitational mass $1.4 M_{\odot}$. }
\begin{ruledtabular}
	\begin{tabular}{cccccccccc}
	EOS&$M$&$M_b$&$M_p$&$-E_t$&$-E_g$&$E_n$&$-E_t/A$&$-E_g/A$&$E_n/A$ \\
           &($M_{\odot}$)&($M_{\odot}$)&($M_{\odot}$)&($M_{\odot}$)&($M_{\odot}$)&($M_{\odot}$)&(MeV)&(MeV)&(MeV)\\
		\hline
          ALF2	&1.40   &1.59 	&1.59 &	0.194 &	0.191 &	-0.003 &	114.373 &	112.423 &	-1.95\\
          \hline
          APR3	&1.40 	&1.56 &	1.62 	&0.160     &0.218 &	0.057 &	96.34 &	130.90 &	34.56\\
          \hline
          APR4	&1.40 	&1.57 &	1.64    &0.170     &0.236 &	0.065 &	101.74&	140.78 &	39.04\\
          \hline
          ENG	&1.40   &1.55 	&1.62   &0.151     &0.218 &	0.067 &	91.56 &	131.93 &	40.37\\
          \hline
          MPA1	&1.40 	&1.55 	&1.61 	&0.145 	&0.208 &   0.064 &88.00 	&126.62 &	38.62\\
          \hline
          SLY	&1.40   &1.55 &	1.63 	&0.145     &0.230 &	0.085 &	88.16 &	139.91 &	51.75\\
          \hline
          WWF1	&1.40 	&1.58 &	1.66 	&0.178     &0.262 &	0.084 &	105.97&	155.63 &	49.67\\
          \hline
          WWF2  &1.40   &1.56 &	1.64 	&0.161 	&0.241 &	0.080 &	96.79 &	145.07 &	48.28\\
          \hline
          QMFL40&1.40 	&1.55 &	1.62 	&0.149 	&0.224 &	0.075 &	90.36 &	135.81& 	45.45\\
          \hline
          QMFL60&1.40 	&1.55 	&1.62 	&0.149 	&0.220 &   0.072 &90.21 	&133.60 &	43.39\\
          \hline
          QMFL80&1.40 	&1.55 &	1.61 	&0.147 	&0.208 &   0.061 &	89.33 &	126.21 &	36.88\\

\hline
CIDDM	&1.40 	&1.70 	&1.58 	&0.301 	&0.176 	&-0.125 	&165.94 	&96.83 	&-69.12\\
\hline
CDDM1	&1.40 	&1.60 	&1.55 	&0.201 	&0.151 	&-0.051 	&117.91 	&88.27 	&-29.64\\
\hline
CDDM2	&1.40 	&1.61 	&1.53 	&0.209 	&0.133 	&-0.076 	&122.24 	&77.71 	&-44.53\\

	\end{tabular}%
	\label{tab:table1}%
 \end{ruledtabular}
\end{table}%

\begin{table}[htbp]
	\centering
	\caption{The binding energies of neutron stars and quark stars with maximum gravitational mass $M_{max}$. }
\begin{ruledtabular}
	\begin{tabular}{cccccccccc}
 EOS&$M$&$M_b$&$M_p$&$-E_t$&$-E_g$&$E_n$&$-E_t/A$&$-E_g/A$&$E_n/A$ \\
        &($M_{\odot}$)&($M_{\odot}$)&($M_{\odot}$)&($M_{\odot}$)&($M_{\odot}$)&($M_{\odot}$)&(MeV)&(MeV)&(MeV)\\
		\hline

ALF2	&2.09 	&2.54 	&2.69 	&0.453 	&0.598 	&0.145 	&167.32 	&221.08 	&53.76\\
\hline
APR3	        &2.39 	&2.96 	&3.37 	&0.567 	&0.972 	&0.405 	&179.89 	&308.18 	&128.30\\
\hline
APR4	        &2.22 	&2.73 	&3.13 	&0.513 	&0.918 	&0.405 	&176.58 	&316.05 	&139.47\\
\hline
ENG	        &2.24 	&2.71 	&3.15 	&0.468 	&0.902 	&0.434 	&162.19 	&312.36 	&150.17\\
\hline
MPA1	    &2.47 	&3.02 	&3.44 	&0.551 	&0.969 	&0.418 	&171.40 	&301.59 	&130.19\\
\hline
SLY	        &2.05 	&2.43 	&2.81 	&0.379 	&0.762 	&0.383 	&146.40 	&294.12 	&147.72\\
\hline
WWF1	&2.14 	&2.64 	&3.04 	&0.507 	&0.904 	&0.398 	&180.00 	&321.17 	&141.17\\
\hline
WWF2	&2.20 	&2.70 	&3.14 	&0.496 	&0.934 	&0.438 	&172.47 	&325.06 	&152.59\\
\hline
QMFL40	    &2.03 	&2.39 	&2.63 	&0.363 	&0.602 	&0.239 	&142.26 	&235.98 	&93.72\\
\hline
QMFL60	    &2.08 	&2.47 	&2.80 	&0.386 	&0.723 	&0.337 	&147.15 	&275.51 	&128.35\\
\hline
QMFL80	&2.11 	&2.51 	&2.81 	&0.401 	&0.708 	&0.307 	&150.06 	&265.05 	&114.99\\
\hline									
CIDDM	&2.07 	&2.64 	&2.58 	&0.570 	&0.514 &	-0.056 	&202.78 	&182.80 	&-19.98\\
\hline
CDDM1	&2.20 	&2.61 	&2.71 	&0.417 	&0.511 	&0.094 	&149.88 	&183.64 	&33.75\\
\hline
CDDM2	&2.41 	&2.87 	&2.99 	&0.453 	&0.572 	&0.119 	&148.35 	&187.45 	&39.10\\

	\end{tabular}%
	\label{tab:table2}%
 \end{ruledtabular}
\end{table}%

\begin{figure}[!htb]
\includegraphics [width=0.7\textwidth]{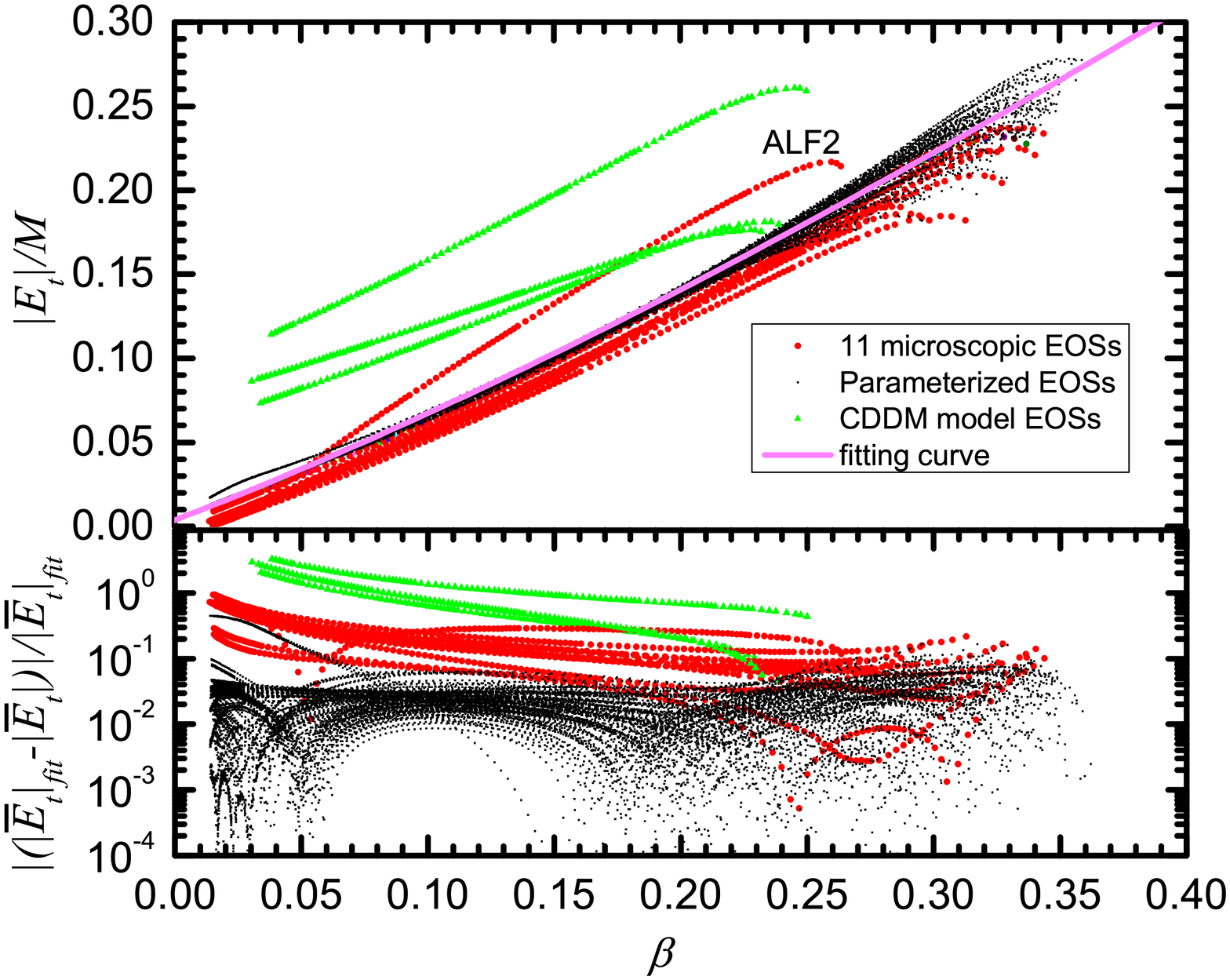}
\caption{ (Top) $|E_{t}|/M-\beta$ relation with various EOSs together with fitting curve (solid curve). (Bottom) The relative fractional difference
between the numerical results and the fitting curve, where $|\bar{E}_{t}|=|E_{t}|/M$. }
\label{Fig.1}
\end{figure}

\begin{figure}[!htb]
\includegraphics [width=0.7\textwidth]{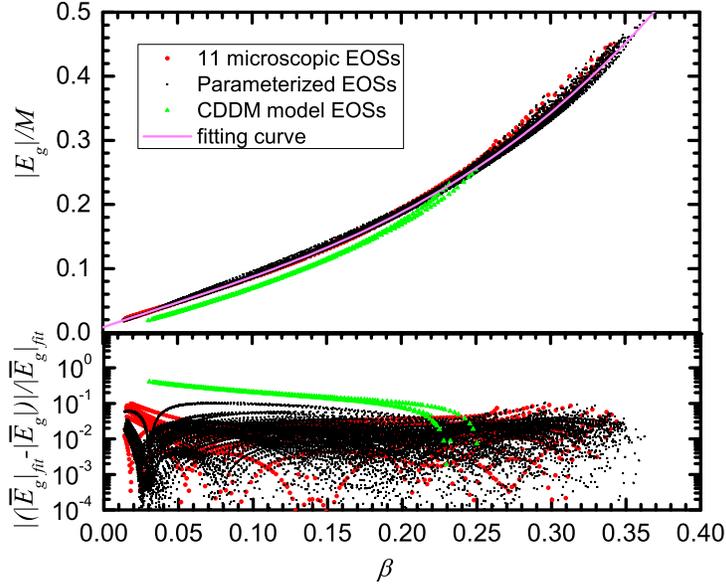}
\caption{ (Top) $|E_{g}|/M-\beta$ relation with various EOSs together with fitting curve (solid curve). (Bottom) The relative fractional difference
between the numerical results and the fitting curve, where $|\bar{E}_{g}|=|E_{g}|/M$.}
\label{Fig.2}
\end{figure}

We first present the universal relation between the  total binding energy and the compactness, as shown in Fig. \ref{Fig.1}. It is shown that for all of the parametric EOSs and ten of the 11 microscopic EOSs (namely, except for the hybrid star model ALF2), there is only a rough universal relation; while for EOSs of quark stars and hybrid stars, their relations are quite divergent. This result further indicates that the total binding energy is not a appropriate parameter to construct the universal relations.
When we replace the total binding energy by the gravitational binding energy to plot the similar relations, we found that there is a much better universal relation for all the parametric EOSs and the 11 microscopic EOSs, as presented in Fig. \ref{Fig.2}. Moreover, all the three quark star EOSs outline a tight-fitting branch in the figure as green dash lines, which is slightly different from the neutron star branch.  Interestingly, the ALF2 EOS, which refers to the typical EOS of hybrid stars, lands on the neutron star branch instead of the quark star branch in Fig. \ref{Fig.2}. These means that the all of the neutron stars with a crust (whatever the composition of the core) have a similar universal relation between the gravitational binding energy and the compactness.

\begin{figure}[!htb]
\includegraphics [width=0.7\textwidth]{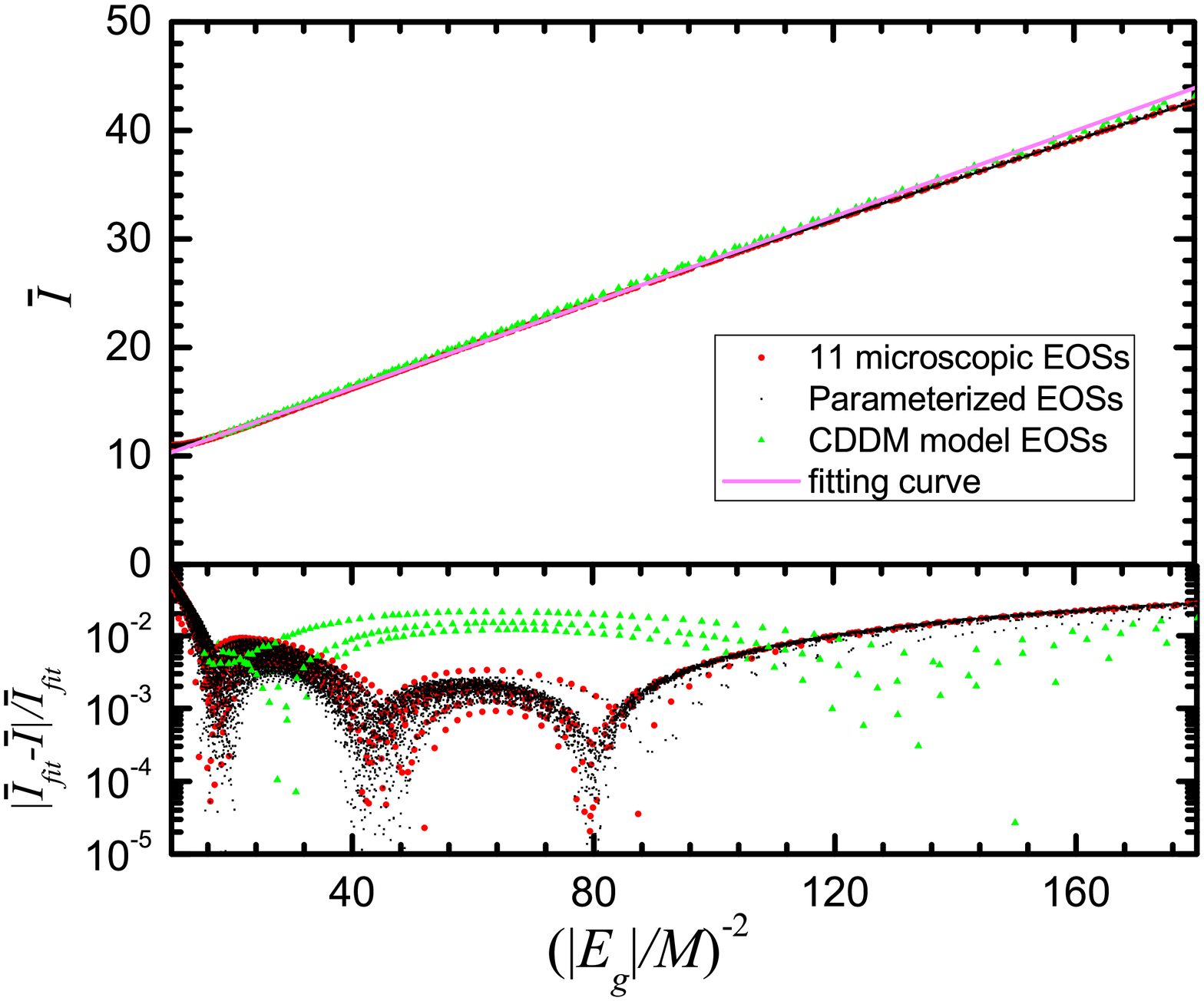}
\caption{(Top) $\bar{I}-(|E_{g}|/M)^{-2}$ relations for various EOSs, together with the fit in Eq. \ref{GI} (solid curve). (Bottom) The relative fractional difference
between the numerical results and the fitting curve. }
\label{Fig.3}
\end{figure}

\begin{figure}[!htb]
\includegraphics [width=0.7\textwidth]{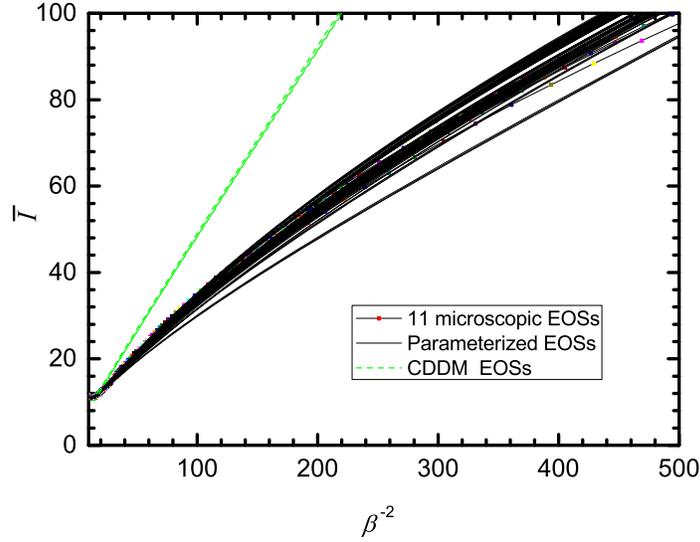}
\caption{Relations between    compactness and   dimensionless moment of inertia. }
\label{Fig.4}
\end{figure}

Inspired by the above results, we expect that there ought to exist some interesting universal relations between the  gravitational binding energy and the global properties of neutron stars. As the moment of inertia is  usually used to investigate the universal relation, here we explore the relation between the moment of inertia and the gravitational binding energy first.
In Fig. \ref{Fig.3}, the universal relation between  the dimensionless gravitational  binding energy  and  the dimensionless moment of inertia is presented. It is shown that there is a perfect linear universal relation between $(|E_{g}|/M)^{-2}$  and $\bar{I}$, which can be
approximated by the formula

\begin{equation}
  \bar{I}=   0.1806(\frac{|E_g|}{M})^{ -2}+  9.314.
\label{GI}
\end{equation}
 Here the neutron stars and quark stars also  follow the same universal relation, as shown in  Fig. \ref{Fig.3}. This  means that we can not distinguish the normal neutron stars and quarks through this relation.
According to Eq. \ref{GI}, it is sufficient to estimate the gravitational binding energy  if the stellar mass $M$ and the moment of inertia $I$ are measured simultaneously in the future, whether the compact star is a quark star or a neutron star.
Optimistically, scientists have  pointed out that
 enough precise measurement of pulsar motion in the double-pulsar system could lead to a relative accurate determination of the moment of inertia of neutron star
\cite{Lyne2004,Lattimer2005,Bejger2005}.
The methods of  applying the universal relations to constrain the relevant quantities also can be found in Ref. \cite{Breu2016}.

For comparison, the relations between the compactness and  the dimensionless moment of inertia are presented in  Fig. \ref{Fig.4}. It is clear that for the adopted EOSs, these relations are rather divergent. From this point of view,  the gravitational binding energy is a better one in expressing the universal relations.

\begin{figure}[!htb]
\includegraphics [width=0.7\textwidth]{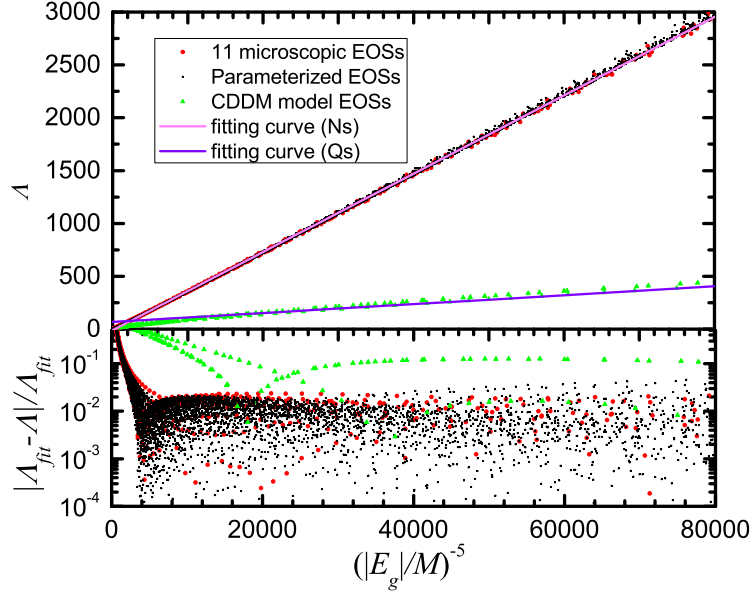}
\caption{(Top) $\Lambda-(|E_{g}|/M)^{-5}$ relations with various EOSs together with fitting curve (solid curve). (Bottom) The relative fractional difference
between the numerical results and the fitting curve.  }
\label{Fig.5}
\end{figure}

\begin{figure}[!htb]
\includegraphics [width=0.7\textwidth]{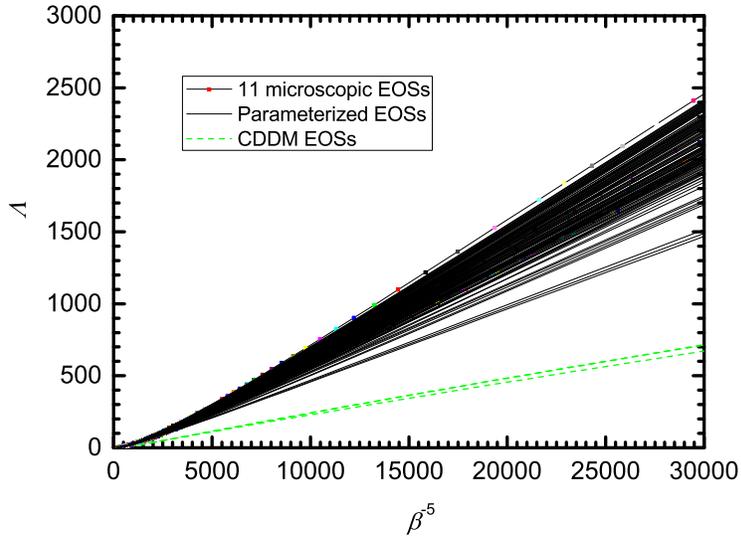}
\caption{Relations between   compactness and  tidal deformability. }
\label{Fig.6}
\end{figure}

With the breakthrough measurement on the gravitational waves in recent years, the tidal deformability is now an important character of a neutron star \cite{LIGO2017}. Many accurate universal relations associated with the tidal deformability (or the tidal Love number)  are proposed, such as the relations of I - Love and Q - Love  \cite{Yagi2013a} and the relations between f-mode frequency/damping time and tidal deformability \cite{Wen2019fmode}. As has discussed above, according to the observed quantities, these relations can be used to learn about the quantity which is inconvenient to be observed.
Currently, the constraint on the tidal deformation from the gravitational-wave detection of GW170817 is $\Lambda_{1.4}=190^{+390}_{-120}$ at  90\% level \cite{LIGO2018}. Interestingly, we found that there exist ideal  linear universal relations between  the negative fifth power of dimensionless gravitational  binding energy $(|E_{g}|/M)^{-5}$ and the tidal deformability $\Lambda$, as shown in  Fig. \ref{Fig.5}, where the normal neutron stars and quark the stars obey two totally different universal relations.
To the normal neutron stars, the universal relation can be approximated by
\begin{equation}
 \Lambda=3.646\times 10^{-2}   (\frac{|E_g|}{M})^{-5} -4.233;
\label{LG1}
\end{equation}
while to the quark stars, its  universal relation can be approximated by
 \begin{equation}
 \Lambda= 3.245 \times 10^{-3}(\frac{|E_g|}{M})^{ -5}  +  108.
\label{LG2}
\end{equation}
Similar to the application of Fig. \ref{Fig.3}, any two of the three quantities ($\Lambda$, $E_{g}$ and $M$ ) are observed precisely, and then we can use the universal relations to constrain the third quantity. Moreover, the different universal relations of the normal neutron stars (with a crust) and the quark stars (without a crust) provide a potential way to distinguish these two kind of compact stars, for example,
 if a relative higher tidal deformability (e.g. $\Lambda>500$) is observed, then we can conclude that it should be a normal neutron star.
 In fact, the different universal relations of these two kind of compact stars may understand through the definition of the tidal deformability,  which is dependent mostly on the internal structure of the star near the outer layer \cite{Hinderer2010}, where normal neutron stars have a completely different outer layer from   quark stars.

Similarly, we also present the relations between  the compactness and the tidal deformability, as shown in  Fig. \ref{Fig.6}.  Through comparing Figs.  \ref{Fig.5} and \ref{Fig.6},
we can see once again that the gravitational binding energy is  better than the compactness in expressing the universal relations.


\begin{figure}[!htb]
\includegraphics [width=0.7\textwidth]{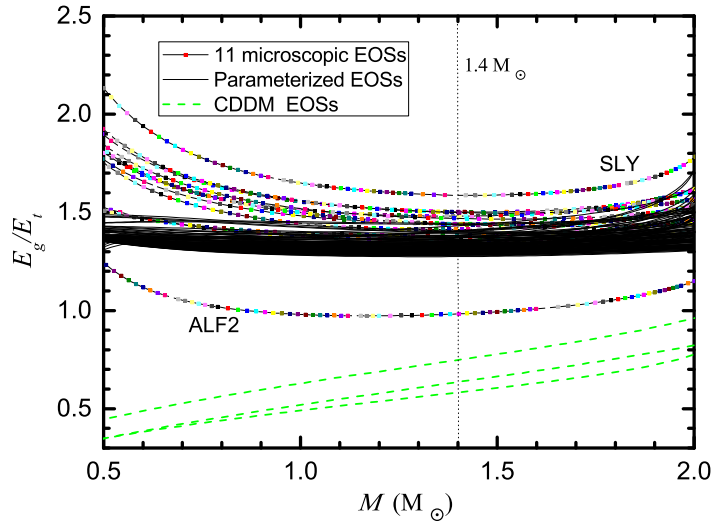}
\caption{The ratio of gravitational binding energy to total binding energy as a function of the stellar mass. }
\label{Fig.7}
\end{figure}

In the end, we would like to give some discussion on the measurement and estimation of binding energy. As has mentioned in Sec. II, through the detection of  supernova explosion, the measurable properties of a neutron star include the effective total binding energy ($E_{et}$), the baryon mass ($M_{b}$) and the gravitational mass ($M$) \cite{Prakash1997,Lattimer2001,Lattimer1989,Lattimer2007,Rosso2017,Hirata1987,Bionta1987,Bethe1995}. Give available and precise enough $E_{et}, M_{b}$ and $M$ from the detection of the supernova explosions in the future, can we estimate the gravitational binding energy? We try to figure out a way to solve this problem.
 As has pointed out above, Table \ref{tab:table1} shows that the gravitational binding energy is about 1.3-1.6 times of the total binding energy for the canonical neutron stars.
 In fact, except for the quark stars and hybrid stars, the ratio of  gravitational binding energy to total binding energy is around 1.3-1.6 for most of the normal neutron stars, as shown in Fig. \ref{Fig.7}. If we believe or can  prove the star is a normal neutron star, then we can roughly estimate the gravitational binding energy.

 In addition, if the compactness of a neutron star can be measured precisely, then the  $E_{g}/M$  can be estimated rather accurately according to the universal relation described in Fig. \ref{Fig.2}, and further the dimensionless moment of inertia and the tidal deformability can be driven from the universal relations in Figs. \ref{Fig.3} and \ref{Fig.5}, respectively.

We also investigated the universal relations in  alternative gravity theory and try to find if the universal relation can be used to distinguish the alternative gravity theory from General Relativity (GR). The Eddington-inspired Born-Infeld (EiBI) theory \cite{Banados2010} is employed as a representation to calculate the universal relations and the results show that there is no distinct difference of the universal relations and we cannot detect the deviations of the alternative gravity theory from GR.

\section{Summary}\label{Summary}
By analyzing the different kinds of  binding energy of neutron star, it is shown that the gravitational binding energy carries the information of the stellar mass distribution and can be used as a parameter to express the universal relations  with the global properties of neutron stars. In this work, two universal relations expressed by the dimensionless gravitational binding energy are presented. They are the universal relation between the gravitational binding energy and the moment of inertia, and the universal relation between the gravitational binding energy and the tidal deformability. It is shown that for the former, both of the normal neutron stars and the quark stars follow the same universal relation; while for the later, the normal neutron stars and quark stars  satisfy  different relations, which can be used to distinguish  a normal neutron star from a quark star.
On the one hand, if any two of the three quantities in the universal relations  are observed precisely,  we can use the universal relations to constrain the third quantity. Thus the universal relations provide a potential way to estimate the gravitational binding energy if the stellar mass and the moment of inertia/the tidal deformability are precisely measured. On the other hand, if the future estimation of the gravitational binding energy through the detection of the supernova explosion or the accurate measurement of the compactness  will benefit us in learning the moment of inertia and the tidal deformability.

It should be noted that the universal relations in this work are obtained based on the  non-rotating neutron star model. These relations  still hold for the neutron stars spinning much slower than the Kepler frequency. However, for a neutron star spinning close to Kepler frequency (such as a newly born neutron star), the universal relations given in this work may not be applicable. In fact, the universal relations among the global properties of the rapidly rotating neutron stars are deeply probed in recent years \cite{Breu2016,Bozzola2018,Gao2019}.
For example, there exist universal relations between the scaled rest mass, gravitational mass and angular momentum for the most rapidly uniformly/differentially rotating neutron stars \cite{Bozzola2018}. It is also found that there is an universal relation between the mass (normalized by the stellar mass of non-rotating and Keplerian rotating neutron stars) and the  spin period (normalized by the Keplerian period) \cite{Gao2019}.  We hope to return to these issues in the near future.

\begin{acknowledgements}
We thank Bao-An Li for helpful discussions. This work is supported by NSFC (Grants No. 11975101 and No. 11722546) and the talent program of South China University of Technology (Grant No. K5180470). This project has made use of NASA’s Astrophysics Data System.

\end{acknowledgements}


\end{document}